\documentstyle[11pt,newpasp,twoside,epsf,psfig]{article} 
\def\edcomment#1{\iffalse\marginpar{\raggedright\sl#1\/}\else\relax\fi} 
\reversemarginpar

\begin{document} 
\title{Probing the physics of relativistic jets through X-ray observations} 
\author{Laura Maraschi \& Fabrizio Tavecchio}
\affil{Osservatorio Astronomico di Brera, Via Brera 28, 20121 Milano, Italy}

\begin{abstract} 
The broad band sensitivity and flexibility of the $Beppo$SAX satellite have
allowed unprecedented studies of the X-ray emission from blazars.  Here
we focus on recent results on the SEDs of a group of blazars with
emission lines, allowing to estimate both the luminosity in the jet and
the luminosity of the accretion disk. Implications for the origin of the
power carried by relativistic jets are considered.
\end{abstract} 

\section{Introduction} 

It is now generally accepted that the blazar "phenomenon" (highly
polarised and rapidly variable radio/optical continuum) is due to a
relativistic jet pointing close to the line of sight (Blandford and Rees,
1978). We further propose that, rather than trying to define a dividing line
 between objects
with or without emission lines (flat spectrum quasars vs. BL Lacs) or
between radio bright and X-ray bright BL Lacs, 
the most productive approach at present is to emphasize the continuity
of properties within a single blazar population.

Here and in the following we will assume that Quasars with Flat
Radio Spectrum (FSQs, which include OVVs and HPQs) and BL Lac objects are
essentially "similar" objects in the sense that the nature of the central
engine is similar apart from some basic scales.  Our aim is to start
 from a (common) physical comprehension of the phenomenology with the goal of 
 understanding the role of more fundamental parameters,  
 like the central black hole mass, angular momentum and accretion rate,
in determining the properties of the jets and of the associated disks.

\section{The Unified Framework for the SEDs of blazars.}
 
It was noted early on that the SEDs of blazars exhibited remarkable
systematic properties (Landau et al. 1986, Sambruna et al. 1996).  The
subsequent discovery by the Compton Gamma Ray Observatory of gamma-ray
emission from blazars (a summary can be found in Mukherjee et al. 1997)
was a major step forward, showing that in many cases the bulk of the
luminosity was emitted in this band and questioning the importance of
previous studies of the SEDs at lower frequencies.

A systematic investigation on the SEDs of the main complete samples of
blazars (X-ray selected, radio-selected and Quasar-like, Fossati et
al. 1998) including gamma-ray data showed that the systematic trends
found previously indeed persisted, suggesting a continuity of spectral
properties (spectral sequence).  All the SEDs show two broad components
with peaks in the $10^{13} -10^{18}$Hz and $10^{21} -10^{25}$Hz ranges
respectively. Both peaks appear to shift to higher frequencies with
decreasing luminosity. We will call red and blue the objects at the
different extremes of the sequence. 

Beamed synchrotron and inverse
Compton emission from a single population of relativistic electrons accounts very well
for the observed SEDs except in the radio to mm range where effects of
selfabsorption and inhomogeneity are important (see also Kubo et al. 1998
). We recall that the relativistic particle spectrum must be "curved" in order to
explain the peaks observed in the SEDs. The curvature is often modelled
with a broken power law. The model predicts that the synchrotron and IC 
emissions should vary in a correlated fashion,
 especially at frequencies near the peaks. Despite
the difficulty of getting adequate data, this has been verified at least
in some well studied objects. In the following we will assume that this
emission model holds in general.

Ghisellini et al. 1998 derived the physical
parameters of jets of different luminosities along the sequence
applying the above model with seed photons of internal (SSC) as well
as external (EC) origin. The results suggest that i) the importance of 
external seed photons increases with
increasing jet luminosity ii) the "critical" energy of the
radiating electrons decreases with increasing (total) radiation energy density.
The latter dependence is physically plausible since the radiation energy
density determines the energy losses of relativistic particles.  If the
critical electron energies were determined by a balance between
injection/acceleration and cooling processes the latter dependence could
be understood.

This "unified" theoretical scheme,  needs to be
tested in many respects. One can think of at least two ways of doing so :
i) determine the physical parameters in individual objects with improved data
ii) understand the mechanisms of particle acceleration and injection from
detailed variability studies.

More indirectly,  if FSQs and BL Lacs
contain "similar" jets (at least close to the nucleus) as suggested by
the continuity of the SEDs, we still need to understand the differences in
emission line properties.  Also in this respect continuity could hold in
the sense that the accretion rate may decrease continuously along the
sequence but the emission properties of the disk may not simply scale
with the accretion rate.

\begin{figure} 
\plotone{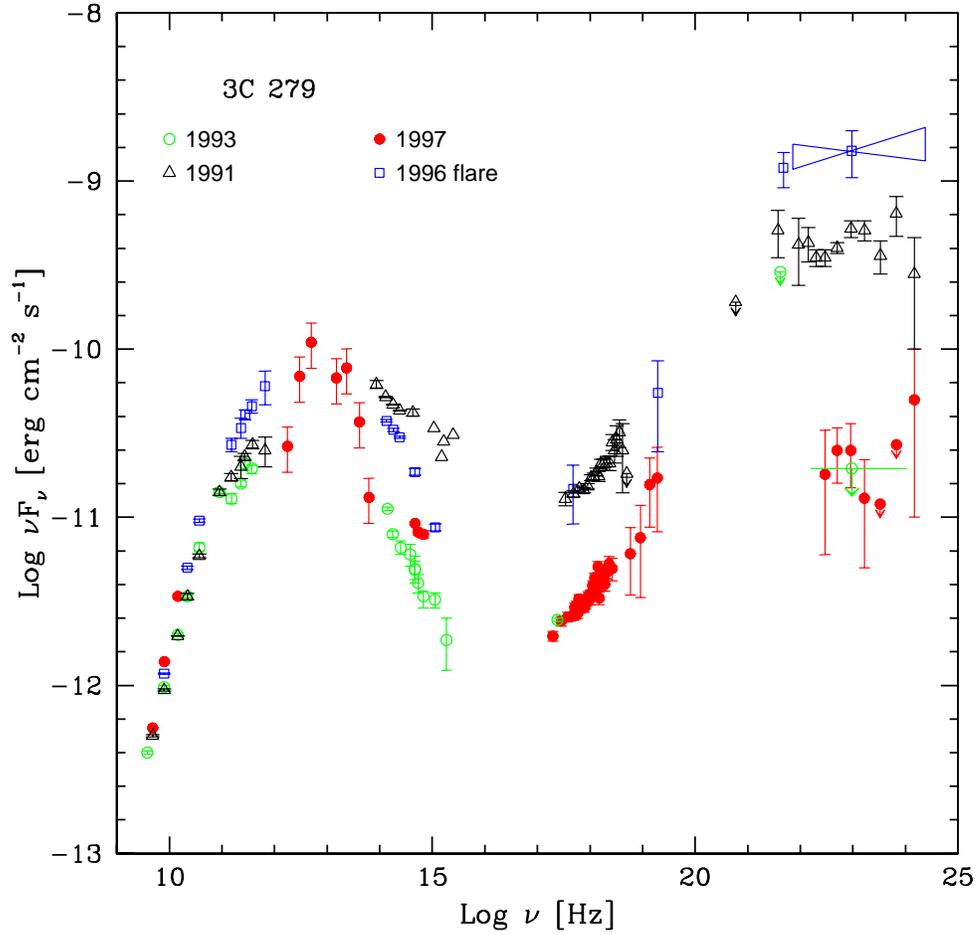}
\caption{Quasi-simultaneous SEDs of the quasar 3C279 taken in the
different epochs. The {\it Beppo}SAX and EGRET data taken in 1997 are
almost exactly contemporaneous, while the ISO spectrum is taken one month
before.}
\end{figure}

\section{Studies of individual objects.}

For red blazars the study of the synchrotron
component is difficult, because  the Synchrotron peak,
falls in the poorly covered IR - FIR range. Furthermore the study
of the gamma-ray component in the MeV-GeV region of the spectrum has been
difficult in the last few years due to the loss of efficiency of EGRET
and is now impossible after reentry of CGRO. Thus relatively little progress 
has been made lately.

We show in Fig. 1 the case of one of the best studied sources,
3C 279, observed with $Beppo$SAX in January
1997 simultaneously with CGRO (Hartman et al. 2000, ) and close in time
(December 1996) with ISO (Haas et al. 1998). The source was found to be in a
rather low state, not far from the low state observed in 1993 (also
shown) (Maraschi et al. in preparation). Also shown for comparison 
are the two highest states recorded in 1991 and 1996l .

A general correlation of the optical X-ray and $\gamma$-ray fluxes is
apparent as predicted. The synchrotron peak however, which would help a lot
constraining models, is difficult to localize.  
The ISO data
are consistent with those at higher and lower frequencies (Jan 1997) at
the two ends of the ISO range but suggest an additional, highly luminous
(of order $10^{46}$ erg s$^{-1}$ thermal component. The origin of such component 
is  not easy to assess.

The situation is better for blue blazars. In several sources of this
class the Synchrotron component peaks in the X-ray band, where numerous
satellites can provide good data. In few bright extreme BLLac objects the
high energy $\gamma$-ray component is observable from ground with TeV
telescopes (for a general account see Catanese \& Weekes 1999). In these
particular cases the contemporaneous X-ray/TeV monotoring demonstrated
well the correlation between the Synchrotron and the IC components. A
very good example is Mkn 421 for which we obtained the observation of a
simultaneous TeV/X-ray flare with Whipple and {\it Beppo}SAX in 1998,
probing for the first time the existence of correlation on short (hour)
time scales (Maraschi et al. 1999; see also Takahashi et al. 1999,
Catanese \& Sambruna 2000). When the position of the two peaks can be well
determined observationally, as is possible in this type of sources,
robust estimates of the physical parameters of the jet can be obtained
(e.g. Tavecchio et al. 1998). This was done for both Mkn 421 and Mkn 501
(Maraschi et al. 1999, Pian et al. 1997, Tavecchio et al., submitted).

The broad band response of $Beppo$SAX allows a reliable determination of
the position of the synchrotron peak, $E_{\rm peak}$, if it falls between
0.1 and 100 keV.  We could verify that during the flare of Mkn 421
mentioned previously $E_{\rm peak}$, moved to higher energies with
increasing intensity (Fossati et al. 2000). The same behaviour was
exhibited in a more dramatic way by Mkn 501. Its synchrotron peak moved
to $E>100$ keV during the extraordinary activity in April 1997 (Pian et
al. 1998). Subsequent snapshot spectra obtained with $Beppo$SAX showed a
systematic decrease of $E_{\rm peak}$ down to $\simeq 0.1$ keV in June
1999 while the source was fading. Over this two year period the X-ray
light curve as measured by the ASM aboard XTE was not monotonic with an
overall decay interrupted by flares. Thus our observations show that
$E_{\rm peak}$ correlates with luminosity not only along individual
flares but also on much longer timescales.

The $E_{\rm peak}$ vs. luminosity relation observed (higher $E_{\rm
peak}$ for higher luminosity) is opposite to that found in the ``spectral
sequence'', where the peak falls at {\it lower} frequencies for objects
of higher luminosity.

One way to reconcile the two types of behaviour is the following.
Suppose that the Lorentz factor of particles emitting at the peak $\gamma
_{\rm p}$ is determined by a balance between cooling and acceleration
processes.

 Given that $t_{\rm cool}={\rm const}/U\gamma $, where $U$ is the total
energy density, and using the general expression $t_{\rm acc}(\gamma
)=\gamma t_{\rm o,acc}$, found in the theory of diffusive shock
acceleration (see e.g. Kirk et al. 1998) one obtains $\gamma _{\rm p}
\propto \left( {Ut_{\rm o,acc}} \right)^{-1/2}$.

This expression is consistent with the correlation found by Ghisellini et
al. (1998), $\gamma _{\rm p}\propto U^{-1/2}$ provided that the basic
acceleration timescale $t_{\rm o,acc}$ is, on average, similar in all
sources.  Flares in single sources can be interpreted as due to the
temporaneous decrease of $t_{\rm o,acc}$ due to changes in the strength
of the acceleration process. The latter scenario seems to apply quite
well to Mkn 501 (Tavecchio et al. 2000, submitted), for which it is
possible to reproduce the observed variability with the only change of
$\gamma _{\rm p}$.

\section{Jet power vs. accretion power}

Finally we wish to report on work in progress on luminous blazars with
emission lines observed with {\it Beppo}SAX. These are at the
high-luminosity end of the sequence, with the Synchrotron peak in the FIR
region. In these sources the X-ray emission is believed to be produced
through the IC scattering between soft photons external to the jet
(produced and/or scattered by the Broad Line Region) and {\it electrons
at the low energy end of their energy distribution}. Thus measuring the
X-ray spectra and adapting a broad band model to their SEDs yields
reliable estimates of the total number of relativistic particles
involved, which is dominated by those at the lowest energies. This is
interesting in view of a determination of the total energy flux along the
jet (e.g. Celotti et al. 1997, Sikora et al. 1997). The "kinetic"
luminosity of the jet can be written as
\begin{equation}
L_{\rm j}=\pi R^2 \beta c \,U \Gamma ^2
\end{equation}
(e.g. Celotti et al. 1997) where $R$ is the jet radius, $\Gamma$ is the
bulk Lorentz factor and $U$ is the total energy density in the jet,
including radiation, magnetic field, relativistic particles and
eventually protons. If one assumes that there is 1 (cold) proton per
relativistic electron the proton contribution is usually dominant.  

 In
high luminosity blazars the UV bump is often directly observed and/or can
be estimated from the measurable emission lines, yielding direct
information on the accretion process in the hypothesis that the UV
emission derives from an accretion disk. Thus the relation between
accretion power and jet power can be explored. This approach was started
by Celotti et al. (1997) but their estimates of $L_{\rm jet}$ were
obtained applying the SSC theory to VLBI radio data which refer to larger
scales.  Our study involves the analysis of the {\it Beppo}SAX data,
modeling the overall SED and deriving the physical parameters for the
jet.  For three sources the results have been already published
(Tavecchio et al. 2000). For other 6 sources the results used here are
preliminary.

As an example of the quality of the data we are using, the SEDs of 2251
+158 and 1641+339 are shown in Fig. 2. The adopted models are also plotted.
Note that 1641+339 was {\it not} detected by EGRET. Nevertheless its
medium to hard X-ray spectrum and overall SED indicate that a gamma-ray
component similar to other sources is likely to be present.

In addition to the  9 emission line blazars recently observed we  included 
in the analysis 3C 279 and four BL Lac objects for which we had 
previuos good quality  {\it Beppo}SAX data (namely BL Lac, ON231, Mkn 501 and
Mkn 421)  (Tagliaferri et al. 2000a,
Tagliaferri et al. 2000b, Tavecchio et al. 2000 submitted, Maraschi et
al. 1999).
 
\begin{figure} 
\hspace{2.0truecm}
\psfig{file=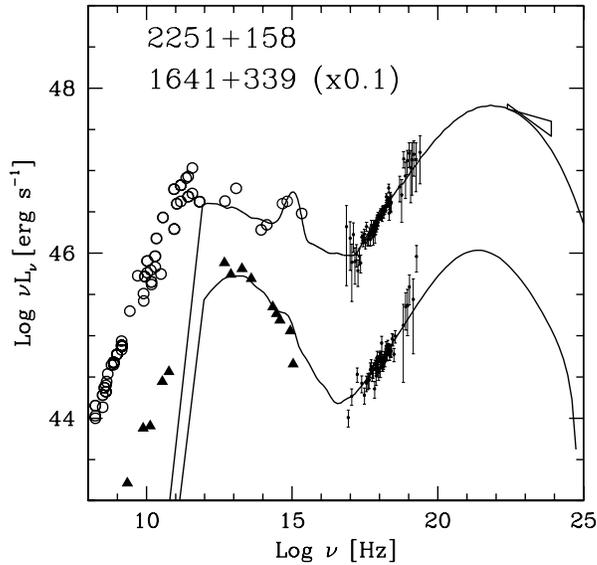,height=9.cm}
\vspace{-0.8truecm}
\caption{Overall SEDs of two emission-line Blazars, 2251+158 (open
circles) and 1641+339 (triangles, for clarity the SEDs has been
multiplied by 0.1). Although the 1641+339 was not detected by EGRET the
overall SED appears similar to those of the other gamma-ray quasars.}
\end{figure}

In all cases physical parameters were estimated by means of a
homogeneous SSC+EC model and the kinetic luminosity of the jet including
protons, $L_{\rm P}$,  as well as the total luminosity radiated by the jet in the
observer frame ($L_{\rm rad}$) were derived accordingly.
The luminosity of the disk could be estimated for all objects except
 the latter three BL Lac, for which 
 we could set only upper limits on the luminosity of their putative
accretion disks. For 3C 279 and BL Lac, the presence of broad
Ly$_{\alpha}$ and H$_{\alpha}$ respectively allowed to estimate the
ionizing continuum (e.g. Corbett et al. 2000).

 The comparison between the total radiative luminosity $L_{\rm rad}$ and
the kinetic luminosity of the jet $L_{\rm P}$ is shown in Fig 3a. The
ratio between these two quantities gives directly the "radiative
efficiency" of the jet, which turns out to be $\eta\simeq 0.1$, though
with large scatter.  The line traces the result of a least-squares fit:
we found a slope $\sim 1$, indicating a rather constant radiative
efficiency along the Blazar sequence (note that the data cover a wide
range of about 5 orders of magnitude).

In Fig. 3b we  compare  the luminosity of the jet, $L_{\rm rad}$,
which is a {\it lower limit} to $L_{\rm jet}$, with the luminosity of 
the disk, $L_{\rm disk}$.

The first important result is that on average the minimal power
transported by the jet is {\it of the same order} as the luminosity
released in the accretion disk. This result poses an important condition
for models elaborated to explain the formation of jets.

Two main lines of approach consider either extraction of rotational
energy from the black hole itself or magnetohydrodynamic winds associated
with the inner regions of accretion disks. Let us parametrize the two
possibilities as follows.  Blandford \& Znajek (1977) summarize the
result of their complex analysis in the well known expression:
\begin{equation}
P_{BZ}\simeq B_0^2 r_g^2 a^2 c  \,\,\,\,\, 
\end{equation}
Assuming maximal rotation for the black hole ($a=1$), the critical
problem is the estimate of the intensity reached by the magnetic field
threading the event horizon, which must be provided by the accreting
matter.  Using a spherical free fall approximation with $B_0^2/ 8\pi \simeq 
\rho c^2$ one can write
\begin{equation}
P_{BZ}\simeq g \dot{M}c^2  \,\,\,\,\, 
\end{equation}
where $P_{acc}=\dot{M}c^2$ is the accretion power and $g$ is of order 1 in 
the spherical case but in fact it is a highly
 uncertain number since it also depends on the field configuration.
  Several authors have recently discussed this
 difficult issue: the arguments discussed by Ghosh \& Abramovicz (1997)
 (GA; see also Livio, Ogilvie \& Pringle 1999) plus equipartition within an
 accretion disk described by the Shakura and Sunyaev (1973) model lead to
 $g \simeq  1$ when gas pressure dominates. Unfortunately at high accretion rates
when radiation pressure dominates its growth is limited.
 Frame dragging by the rotating hole may however increase $g$ to
 values even larger than 1 (Meier 1999). 

 As argued strongly by Livio et
 al. the accretion flow itself may power jets through a hydromagnetic
 wind. However for consistency only some fraction $f \dot{M}c^2$ can be
 used to power the jet.  Further recall that the luminosities observed
 from the jet and disk are related to their respective powers by
 efficiency factors $L_{rad}=\eta P_{jet}$ ; $ L_{disk}=\epsilon P_{acc}$.

Using the condition that $P_{jet} \leq (P_{BZ} + f P_{acc} )$
together with the previous relations we finally find
\begin{equation} 
L_{rad} \leq \frac{\eta (g+f)} {\epsilon }L_{disk}.
\end{equation}
One can account for the observed relation between $L_{rad}$ and $
L_{disk}$ if for instance $\eta\simeq \epsilon \simeq 0.1$ and $f$ or $g$ or
both are close to 1. It is also possible that $\epsilon << 0.1$ if the
optically thick emission derives from radii larger than $3 r_G$ or if the
disk is optically thin as may happen at low accretion rates (Blandford 1990).
This may  be necessary at low luminosities to explain why disks are not seen
(upper limits).  However at high luminosities there is an additional
condition deriving from the total luminosities observed. At $10^{47}$ erg
s$^{-1}$ the Eddington luminosity implies a mass of $10^9$ $M_{\odot}$ and
an accretion rate of $10$ $M_{\odot}$ y$^{-1}$ for $\epsilon = 10^{-1}$. It
seems then difficult to invoke lower efficiencies.

For comparison we show in Fig 3b the estimates for the rotational power
derived by GA for various values of the mass of the central black hole as
a function of the luminosity observed from the disk. The latter is
related to the accretion rate which appears in the formulae of GA
adopting $\epsilon \simeq 0.1$.  Clearly the model fails to explain the
large power observed in the jets of bright quasars, even for BH masses
($M\sim 10^9 M_{\odot}$).  This is because in the radiation pressure
dominated region of the disk the pressure used to estimate the magnetic
field via equipartition does not increase with the accretion rate.

\begin{figure} 
\plottwo{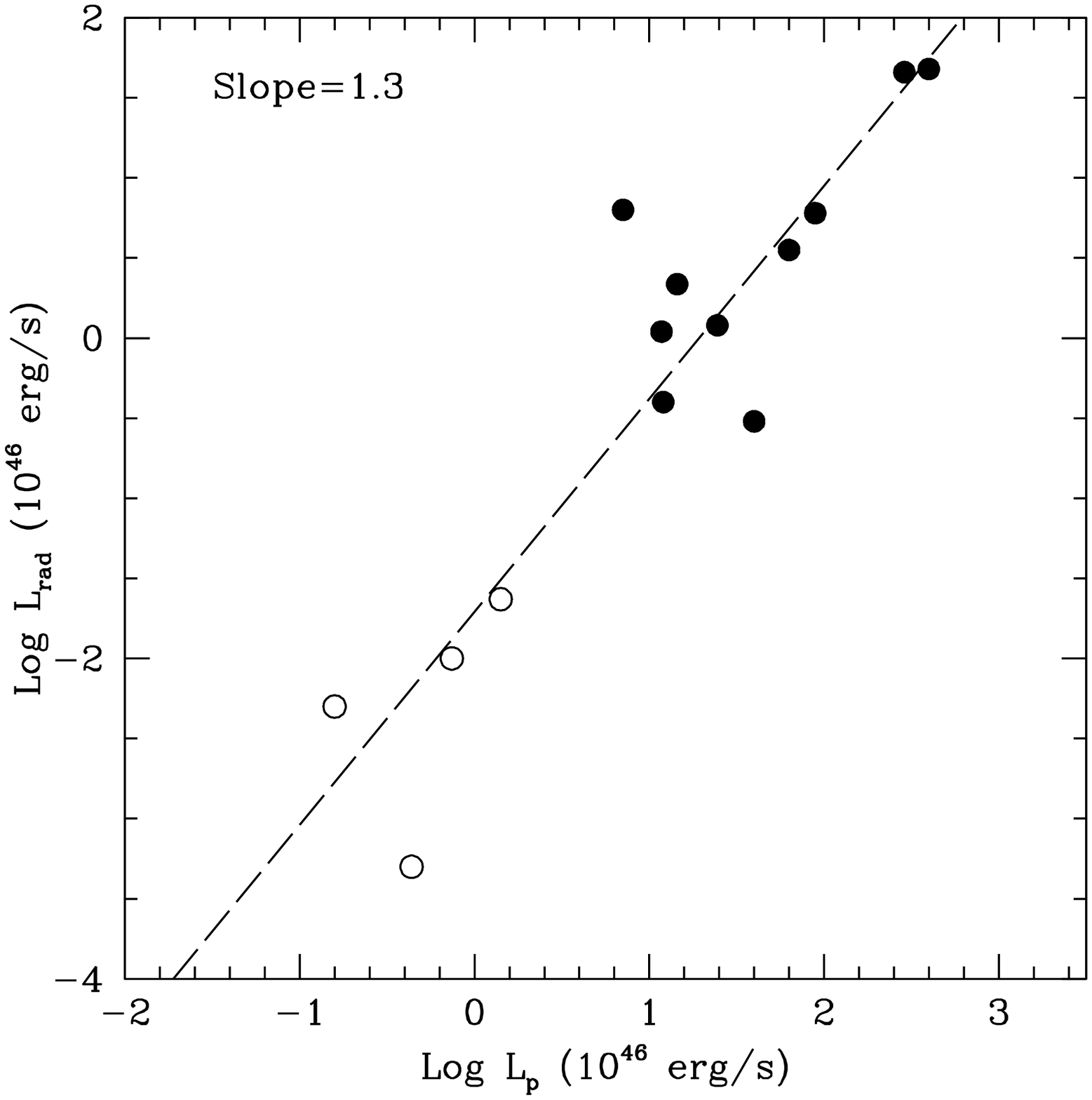}{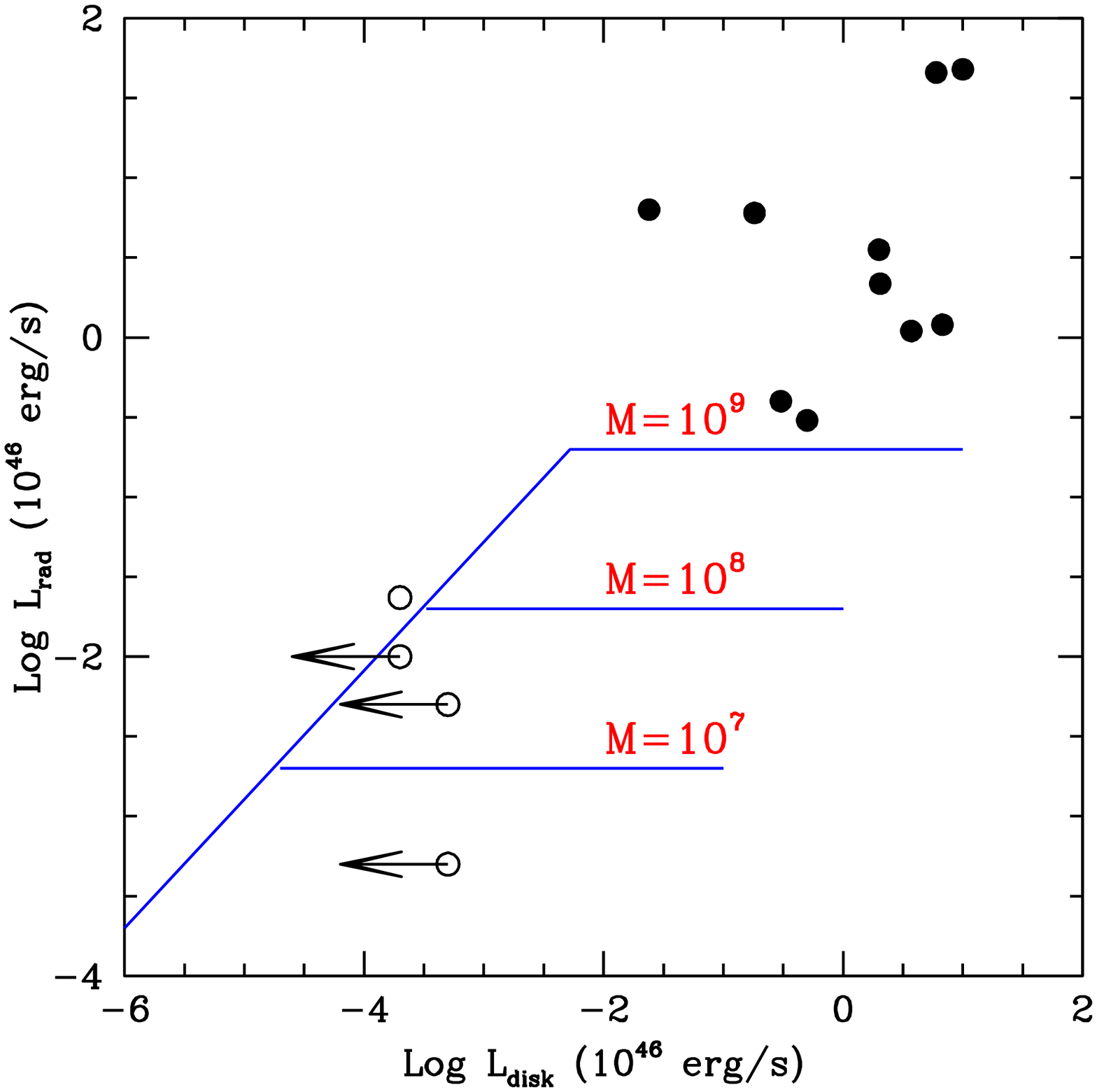}
\caption{{\it Left:} Radiative luminosity vs. jet power for the sample of
Blazars discussed in the text (open circles represent BL Lac
objects). The dashed line indicates the least-squares fit to the
data. {\it Right:} Radiative luminosity of jets vs disk luminosity. The
solid lines represent the {\it maximum} jet power estimated for the
Blandford \& Znajek model for black holes with different masses (in 
Solar units).}
\end{figure}

\section{Conclusions}
The study of broad band SEDs and their variability is essential for
understanding blazars. A unified approach is possible and should be tested.
  While the phenomenological
framework is suggested to be "simple" (e.g. "red" blazars are highly
luminous, have low average electron energies and emit GeV gamma-rays 
while "blue" blazars have low luminosity, high average electron energies
and emit TeV gamma-rays) we do not yet know what determines the emission
properties of jets of different power nor what determines the jet power
in a given AGN. There is however the exciting prospect that such
problems can be tackled with data that can be gathered in the near
future.

\section{References}

Blandford, R.D., \& Znajek, R.L., 1977, MNRAS, 179, 433\\
Blandford, R.D., \& Rees, M.J., 1978, Pittsburgh Conf. on BL Lac
Objects, p. 341-347.\\
Blandford, R.~D., 1990, in ``Active galactic Nuclei'', Springer-Verlag\\ 
Catanese, M., \& Weekes, T.C., 1999, PASP, 111, 1193\\
Catanese, M.\& Sambruna, R.\ M.\ 2000, ApJ, 534, L39\\ 
Celotti, A., Padovani, P., \& Ghisellini, G. 1997, MNRAS, 286, 415\\ 
Corbett, E. A., et al. 2000, MNRAS, 311, 485\\ 
Fossati, G., et al. 1998, MNRAS, 299, 433\\
Fossati, G., et al. 2000, ApJ, 541, 166\\ 
Ghisellini, G., et al. 1998, MNRAS, 301, 451\\
Ghosh, P., \& Abramowicz, M.A. 1997, MNRAS, 292, 887\\ 
Haas, M., et al. 1998, ApJ, 503, L109\\
Kirk, J.G., Rieger, F.M., \& Mastichiadis, A., 1998, A\&A, 333, 452\\
Kubo, H., et al. 1998, ApJ, 504, 693\\
Landau, R., et al. 1986, ApJ, 308, 78\\
Livio, M., Ogilvie, G. I. \& Pringle, J. E. 1999, ApJ, 512, 100\\
Maraschi, L., et al. 1999, ApJ, 526, L81 \\
Meier, D. L. 1999, ApJ, 522, 753\\
Mukherjee, R., et al. 1997, ApJ, 490, 116\\ 
Pian, E., et al. 1998, ApJ, 492, L17\\
Sambruna, R.,M., Maraschi, L., \&Urry, C.M., 1996, ApJ, 463, 444\\
Shakura, N. I. \& Sunyaev, R. A. 1973, A\&A, 24, 337\\ 
Sikora, M., et al. 1997, ApJ, 484, 108\\  
Takahashi, T., et al. 1999, Astroparticle Physics, 11, 177\\ 
Tavecchio, F., Maraschi, L., \& Ghisellini, G., 1998, ApJ, 509, 608\\ 
Tavecchio, F., et al. 2000, ApJ, 543, 535\\

\end{document}